\documentclass[]{spie}  %>>> use for US letter paper
\usepackage{gensymb}
\usepackage{amsmath,amsfonts,amssymb}
\usepackage{graphicx}
\usepackage[colorlinks=true, allcolors=blue]{hyperref}
\usepackage{soul}
\usepackage{booktabs}
\usepackage{tabularx}

\newcommand{\arcsec}{$^{\prime\prime}$}
\newcommand{\ie}{\textit{i.e.}}

\title{METIS high-contrast imaging simulations: \\
From instrument modelling to science readiness}

\author[a]{Gilles Orban de Xivry}
\author[a]{Olivier Absil}
\author[b]{Iain Hammond}
\author[b]{Thomas Bertram}
\author[b]{Roy van Boekel}
\author[b]{André Boné}
\author[b]{Ga\"el Chauvin}
\author[c, d, a]{Valentin Christiaens}
\author[e]{Denis Dolkens}
\author[f]{Gilles Otten}
\author[g]{Prashant Pathak}
\author[h]{Nu\~no Pereira}

\affil[a]{Space sciences, Technologies, and Astrophysics Research (STAR) Institute, Universit\'e de  Li\`ege, all\'ee du Six Ao\^ut 19c, 4000 Liège, Belgium}
\affil[b]{Max-Planck-Institut f\"ur Astronomie, K\"onigstuhl 17, 69117 Heidelberg, Germany}
\affil[c]{Université Paris-Saclay, Université Paris Cité, CEA, CNRS, AIM, F-91191 Gif-sur-Yvette, France}
\affil[d]{Universit\'e Paris-Saclay, CNRS, Institut d’Astrophysique Spatiale, 91405 Orsay, France}
\affil[e]{NOVA Optical and Infrared Instrumentation Group, P.O. Box 2, 7990 AA Dwingeloo, The Netherlands}
\affil[f]{Academia Sinica Institute of Astronomy and Astrophysics (ASIAA), 11F Roosevelt Rd, 10617 Taipei, Taiwan}
\affil[g]{Department of SPASE, Indian Institutes of Technology, Kanpur 208016, Uttar Pradesh, India}
\affil[h]{Royal Belgian Institute for Space Aeronomy (BIRA-IASB), Avenue Circulaire 3, 1180 Brussels, Belgium}

\authorinfo{Further author information: (Send correspondence to G. O. X.)\\G. Orban de Xivry: E-mail: gorban@uliege.be}

\begin{document} 
\maketitle

\begin{abstract}
The Mid-infrared Extremely Large Telescope (ELT) Imager and Spectrograph (METIS) instrument, expected to see first light in early 2030, aims to detect and characterise exoplanets and circumstellar disks through high-contrast imaging (HCI) and spectroscopy. The High-contrast End-to-End Performance Simulator (HEEPS), initially developed to support the design of the METIS HCI modes, has evolved into a crucial tool for the METIS science team to prepare and optimize observations.
HEEPS is an open-source Python-based software with a modular architecture, integrating the wavefront Fresnel propagation package PROPER, and HCI image processing with the Vortex Image Processing (VIP) package. 
%\hl{It can also be used jointly with the telescope observation simulator, ScopeSim. 
Though designed for METIS, its modularity has been applied to other HCI instruments as well.
This work presents recent updates to HEEPS, including modelling of the final METIS pupil and Lyot stops, revised quasi-static non-common path aberrations (NCPA) and Talbot effect simulations informed by as-built optical surface errors, and updated METIS Single Conjugated Adaptive Optics (SCAO) simulations. We also discuss advancements in NCPA control strategies focusing on framerate, latency and sensing performance optimization, particularly for mitigating water vapor seeing effects using the asymmetric Lyot wavefront sensor (ALF) algorithm. 
With these refinements, we present a comprehensive grid of HCI performance simulations for METIS, covering a range of magnitudes in the $L$, $M$, and $N$-bands, and several HCI observing modes. These simulations produce updated 5-sigma sensitivity contrast curves and mock HCI observations, providing key insights on HCI performance for instrument optimization and science observation planning. Our results underscore the key role of end-to-end simulations in bridging instrumental design and scientific readiness in the ELT era.

\end{abstract}


\section{INTRODUCTION}
\label{sec:intro}  % \label{} allows reference to this section
The Mid-infrared ELT imager and spectrograph (METIS) is a first generation instrument of the Extremely Large Telescope currently under construction at Cerro Armazones. 
% It will provide L, M, N band imaging and high dispersion spectroscopy, with high-contrast capabilities. 
Its is expected to see first light in early 2030\cite{Brandl+24,Brandl+26}.
% At the core of its science goals is the detection and characterization of exoplanets and protoplanetary disks. This is enabled by the METIS cryogenic instrument featuring a high-performance adaptive optics module and advanced coronagraphic concepts that can be combined with L-, M-, or N-band imaging (3-13$\mu$m) or high-dispersion spectroscopy (R=100,000) from 3 to 5 $\mu$m.
METIS is a cryogenic instrument, featuring a high-performance adaptive optics (AO) module and advanced coronagraphic concepts that can be combined with $L$, $M$, or $N$-band imaging (3 - 13$\mu$m) or high-dispersion integral field spectroscopy (R=100,000) from 3 to 5 $\mu$m.

% advanced coronagraphic concept
To take full advantage of the exquisite angular resolution and sensitivity  delivered by the ELT and METIS, among the different coronagraphic modes, two flavors of vortex coronagraph are implemented in METIS to reach the smallest inner working angles ($\sim$ 1 $\lambda/D$) while maintaining a high throughput: the classical vortex coronagraph (CVC) and the ring-apodized vortex coronagraph (RAVC).
The CVC mode is based on a vortex phase mask in the focal plane combined with a downstream Lyot stop, and provide the highest throughput. The RAVC, implemented for $L$ and $M$-bands only, uses an additional ring apodizer in the upstream pupil plane to optimize the performance in the speckle-dominated regime.

Those key features are expected to bring major and exciting breakthroughs in exoplanet science, from the detection and characterization of low-mass planets (sub-Neptunes down to rocky planets) to the observation of protoplanets in the early stages of formation, and the study of  disk structures leading to the formation of extrasolar planets. 

Prosaically, the METIS high-contrast imaging (HCI) requirement defined in $L$-band is to reach a post-processed 5-sigma sensitivity of 3 $\times$ 10$^{-5}$ at an angular separation of 5 $\lambda / D$, or $\sim$ 0.1\arcsec , on a relatively bright star ($L$ $\leq$ 6) with a 1-hr observing sequence. Since the requirement pertains to post-processed contrast, it can only be verified by end-to-end simulation. This is the original motivation for developing the High-contrast End-to-End Performance Simulator (HEEPS): support the design of the METIS HCI modes to ensure that the top-level requirement can be fulfilled.
As the design and manufacturing is well underway, and the METIS science team is gearing up, HEEPS is now becoming a crucial tool to prepare and optimize future observations.

In this proceeding, we provide an update on the modelling of various effects affecting the HCI performance, integrating the latest design and  analyses to our end-to-end simulations. We then present our current activities supporting the METIS science preparation, namely the generation of a comprehensive grid of HCI simulations. The two main outputs are contrast curves, which can be used to estimate yield, and mock observations, which enable studying the detectability of more complex objects than just point-like sources, such as extrasolar systems.

% \newpage
\section{HEEPS SIMULATION PIPELINE}\label{sec:heeps}
HEEPS, the High-contrast ELT End-to-end Performance Simulator, has been developed to evaluate  HCI performance and verify  post-processed contrast requirements.
Although it is designed for METIS, its modularity has been applied to other HCI instrument, see Ref.~\citenum{Miller+25} for a recent example.

\begin{figure}[h!]
    \centering
    \includegraphics[width=1\linewidth]{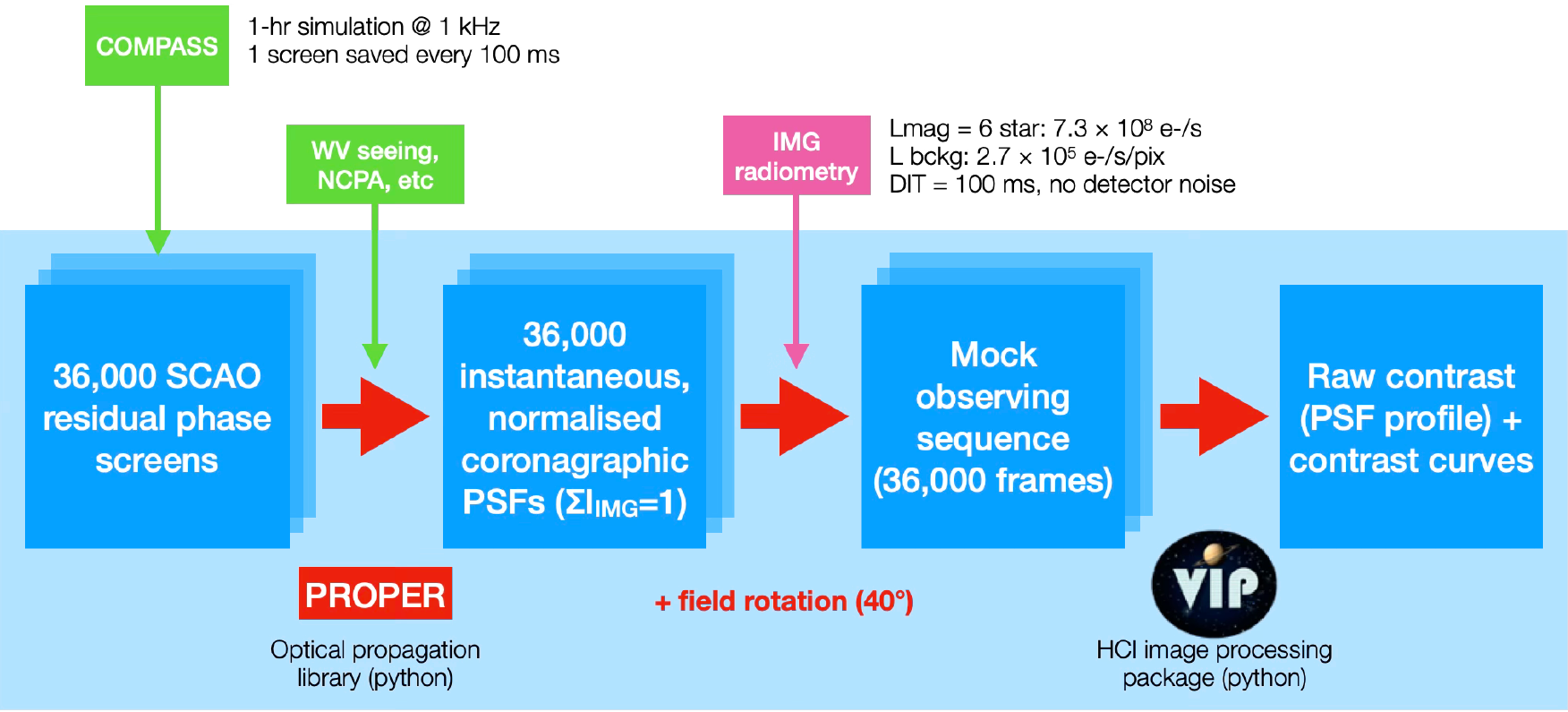}
    \caption{The HEEPS simulation pipeline. HEEPS takes AO phase screens as input, typically simulated with COMPASS, propagate them through an instrument model using PROPER, produce a mock observing sequence, and post-process the frames using the VIP pacakge to produce contrast curves.}
    \label{fig:heeps}
\end{figure}

The HEEPS pipeline is illustrated in Fig. \ref{fig:heeps} and involves the following steps: 
\begin{enumerate}
    \item  Obtain a temporal series of adaptive optics residual phase screens from an end-to-end AO simulation tool.  For METIS, the single-conjugate adaptive optics (SCAO) is simulated with a dedicated extension of the COMPASS\cite{Gratadour+16,Feldt+24} AO simulation tool.
    %end-to-end a 1-hr SCAO simulation with COMPASS to obtain residual phase screens at a 300ms sampling, 
    \item After adding any relevant environmental or instrumental effects (e.g., non-common path aberrations), propagate the residual phase screens through an optical model of the METIS HCI system, using the PROPER\cite{Krist07} software package, to produce a time series of instantaneous coronagraphic PSFs.
    \item With the instantaneous coronagraphic PSF series, produce a mock Angular Differential Imaging (ADI)\cite{Marois+2006} observing sequence.
    \item Compute the classical ADI  post-processed contrast (which includes algorithmic throughput and small-samples statistics), using the VIP\cite{GomezGonzalez2017, Christiaens+23} software package.
\end{enumerate}

%More details on the HEEPS pipeline and our end-to-end simulations can be found in [17].}

By default, HEEPS assumes a Cerro Armazones latitude of -24.59\degree~ and a target declination of -5\degree, which provides a parallactic angle rotation of about 40\degree~in a 1-hr ADI observing sequence.
The typical simulation features a total duration of 1-hr with phase screens sampled every 100 ms, giving a total of 36,000 phase screens to propagate. The photometry of each individual PSF image is thus scaled to a detector integration time of 100ms.

% Effect of down-sampling from 100 to 300ms and more has been studied by Carlomagno...

% \newpage
\section{UPDATES AND DRIVERS OF HCI PERFORMANCE}

% *** Revised assumptions
% - Lyot stop
% - new Chromatic beam wander and amplitude errors
% - New wavefront error induced by water vapor seeing
% - New SCAO simulations... (which specific feature to highlight?)
% - New analyses on focal plane wavefront sensing with ALF

As the design and manufacturing progress, we can refine our analyses and update the modelling of the METIS instrument and of the various effects affecting the HCI performance.

The HCI performance is influenced by a wide range of instrumental and environmental effects. This includes effects seen by the SCAO, such as atmospheric turbulence conditions, misaligned segments in the ELT-M1, non-uniform segment reflectivity, or pupil stability, but also unseen by the SCAO such as chromatic leakage in the coronagraphic mask, pointing drifts and non-common path aberrations, and amplitude effects. The individual and combined effects of each contributor have been the subject of detailed analyses\cite{Carlomagno+20,Delacroix+22}. 

\paragraph{Update to the instrument model.} The key update in our instrument model consists of incorporating the final designs of the METIS pupil stops. 
The prime goal of the  pupil stops within the METIS
cryostat (also referred to as cold stops) is to prevent external background (originating from outside the telescope pupil) from making it to the science detectors, taking into account the contribution due to pupil blurring.
In the context of vortex coronagraphy, the cold stop also acts as Lyot stop. It blocks the diffracted stellar light by the vortex focal plane mask in the downstream pupil plane.
%The vortex focal plane mask redistributes the light outside of the geometric image of the telescope pupil in the downstream pupil plane. This diffracted stellar light is blocked by the pupils stop, also referred to as Lyot stop.
%
This is referred to as the classical vortex coronagraph (CVC), where the star light rejection is only perfect for a circular, unobstructed pupil, and is degraded for non-circular pupil geometries. The effect of the central obscuration, which adds a considerable amount of stellar light within the geometric image of the pupil, can be mitigated by introducing a
greyscale apodizer in an upstream pupil plane, leading to the ring-apodized vortex coronagraph (RAVC) concept. With the RAVC, the level of light inside an annular region of the downstream pupil is strongly reduced, in principle down to zero for a purely annular pupil.
The design of the cold stops is thus a delicate balance between the background (due to misalignment or pupil blurring), the amount of stellar light making it through the Lyot stop, and the throughput of the stop.
Finally, some of the pupil stops need to be asymmetric to enable instantaneous focal-plane wavefront sensing using the ALF deep learning framework\cite{OrbandeXivry+24}, a topic discussed in section \ref{sec:fpwfs}. 

The METIS pupil stop designs balance those various considerations, providing aggressive and conservative strategies. The aggressive strategy does not consider pupil blurring and is driven by the lessons learned from the on-sky operations of NACO and ERIS at the VLT, where it was discovered that maximizing the throughput should generally be promoted over a perfect stopping of the thermal background coming from the telescope structure. On the other hand, the conservative strategy includes additional margins and larger asymmetries for more conservative thermal background management and better focal-plane wavefront sensitivity.

\begin{figure}
    \centering
    \includegraphics[width=0.8\linewidth]{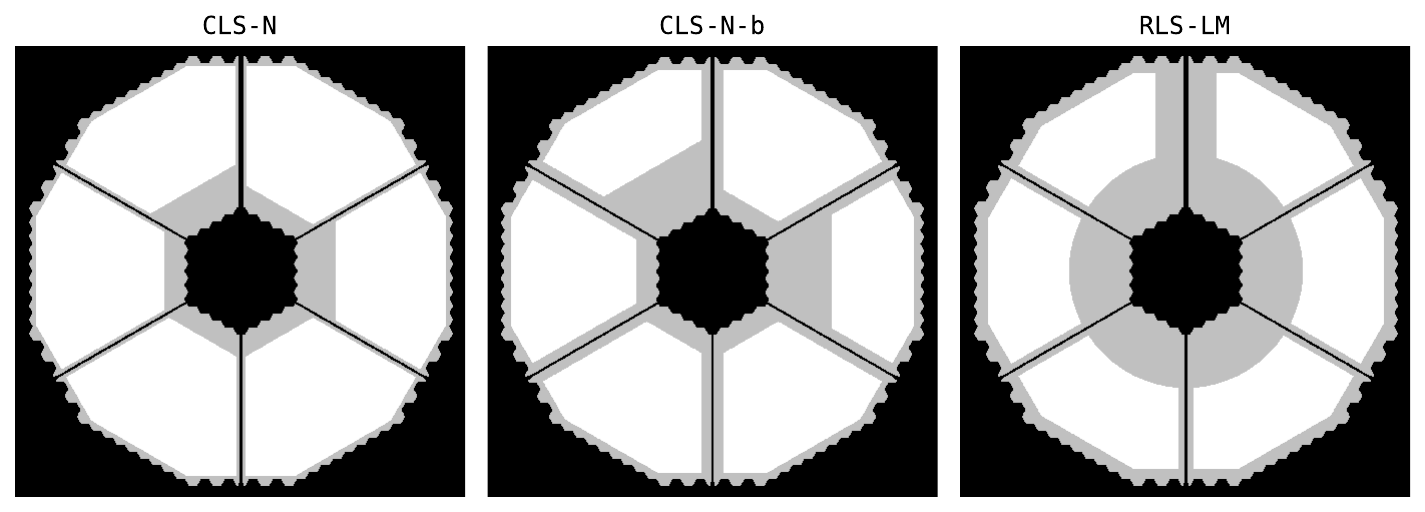}
    \caption{Illustration of three METIS Lyot stops (grey) overlaid on the ELT entrance pupil (black). (Left) the baseline (`aggressive') $N$-band CVC Lyot stop, (Middle) the `conservative' $N$-band CVC Lyot stop, (Right) the $LM$-band Lyot stop for the RAVC. }
    \label{fig:lyot_stops}
\end{figure}

In Fig. \ref{fig:lyot_stops}, we illustrate three different cold stops super-imposed on the entrance ELT pupil: the aggressive and conservative Lyot stops for the $N$-band, and the Lyot stop for the $LM$-band RAVC. The asymmetric pattern and its footprint have been optimized to maximize wavefront sensitivity based on the focal plane images while minimizing throughput loss. The conservative design has a larger asymmetry building more margin for wavefront sensing sensitivity at the expense of throughput. The RAVC stop has a larger central obscuration which is set by the design of the upstream apodizer to block the redistributed light.

\subsection{Drivers of HCI performance}
METIS HCI performance is influenced by a wide range of instrumental and environmental effects, all of which can be enabled or disabled in our simulations. 
Here, we describe which effects are included or excluded in our new analyses and simulation grid, explain the rationale behind these selections, and revisit the prescription for each.

\subsubsection{Effects not included}
% In this update, we can restrict the range of simulated effects to the atmospheric turbulence residuals, non-common path aberrations (see below) and amplitude effects. 

Our decision to exclude certain  effects is based on previous analyses and measurements (e.g., Refs.~\citenum{Delacroix+22, Delacroix+24}) and aims to focus on the key performance drivers while reducing computational complexity. 

Specifically, we exclude the following effects:
\begin{itemize}
\item \textbf{Chromatic leakage of the coronagraphic mask:}  Since the masks are now manufactured and characterized, their excellent performance and low chromatic leakage have been confirmed\cite{Delacroix+24}. At these levels, the residual leakage has a negligible impact on  contrast\cite{Delacroix+22}.
\item \textbf{ELT-M1 primary mirror segment reflectivity:}  The non-uniform pupil amplitude due to segment-to-segment reflectivity variations  has been shown to have a negligible impact on contrast and can be safely  ignored in our subsequent analyses.
\item \textbf{ELT-M1 pupil shape and missing segments:} The influence of the pupil shape is  critical and the ELT primary mirror (M1) is not coronagraphy-friendly due to its large central obscuration, support structure, and segmented design. On top of the baseline shape of ELT-M1, the regular cleaning and re-coating of the M1 segments means that segments could be occasionally misaligned, with up to seven misaligned segments arranged in a flower pattern. The impact on HCI performance was studied in Ref.~\citenum{Delacroix+22}, showing the detrimental effect of misaligned segments. Considering the expected occasional nature of such event and its large impact, we do not include this effect assuming that HCI observation will only be performed with a fully co-phased ELT primary mirror. 
\item \textbf{Pupil stability \& drift:} The ELT pupil stabilization mechanism will operate downstream of the first pupil plane, where the ring apodizer (RAP) is installed. Although the RAVC Lyot stop includes margin to account for such shifts, worst-case  drifts on the same timescale as the ADI observing sequence could produce PSF variations not fully captured by  median ADI post-processing. %This effect should only affect the RAVC observing mode. 
Since the RAVC mode will only be used for obtaining the best contrast in the speckle-limited regime on bright sources,  it will also require the best performance from the telescope and the instrument to effectively deliver superior performance. Hence we decide to not include this effect, assuming that we will only use this mode under the most stable conditions.
% Hence the RAP could be shifting with respect to the ELT, and this slow drift could cause significant 
\end{itemize}

% Overall, NCPAs play a dominant role, and our ability to measure and correct them by offsetting the AO will be a key driver for the final HCI performance. In Section 2.5, we presented the expected sources of NCPA. Here, we will discuss our mitigation strategies in detail in Section 7.3.1. In Section 7.4, we summarize the expected HCI performance.

\subsubsection{Effects included} 
The key drivers of the HCI performance are phase and amplitude errors. Indeed, beyond AO phase residuals, quasi-static and dynamic non-common path aberrations (NCPAs) strongly impact post-processed contrast. Finally, amplitude errors due to Talbot effect can also have a negative impact on contrast at mid-separations (between the speckle and background-limited regime).

%We describe here the effects that we are actually including in our HCI performance analysis, and how they are simulated.

%Beyond AO residual phase screens, quasi-static and dynamic NCPAs are key drivers of the final HCI performance. There two main contributions to varying NCPA in METIS: 

Specifically, the phase and amplitude contributors are the following:
\begin{itemize}
\item \textbf{Adaptive optics residuals:}
The Single-Conjugated Adaptive Optics system\cite{Bertram+26,Feldt+24} is simulated with the COMPASS platform, extended specifically for METIS \cite{Gratadour+16, Feldt+24}.
This extension includes the METIS reconstruction scheme based on a virtual DM and regularised MMSE inversion and projection on the M4 modes. It also includes a simulator of the ELT M4 Central Control System. More details can be found in Ref.~\citenum{Feldt+24}.
Using the COMPASS-METIS platform, we simulate 1-hr  AO closed-loop sequence, saving residual phase screens every 100~ms, which results in 36,000 phase screens.
\item \textbf{Chromatic beam wander (CBW) phase and amplitude errors:} The differential atmospheric refraction between the AO operating wavelength and the scientific camera imparts a slight offset to the beams as they propagate through the atmosphere and the optics. As the telescope tracks the target star, the scientific beam slowly wanders on the optics picking differential aberrations. A detailed study of this effect has been presented in Ref.~\citenum{Bone+26} based on a complete model of the METIS optical train. Using the latest design and incorporating measured surface form errors when available, we have generated new and more realistic variable NCPA time series and included them in our HCI performance analysis. Finally, in addition to creating variable phase aberrations, chromatic beam wander also leads to variable amplitude aberrations that develop upon propagation. Those amplitude aberrations are modeled in parallel to the phase aberrations induced by chromatic beam wander, as described above. See also Refs.~\citenum{Bone+20,Bone+26}.
\item \textbf{Water vapour seeing (WV):} This dynamic contribution of NCPA is due to the non-uniform distribution of water vapor and its strong chromaticity at infrared wavelengths. The non-uniformity generates spatial inhomogeneities in the optical path that are transported by the wind following the frozen flow hypothesis. The water vapor induced phase errors results from the strongly chromatic refractive index exhibiting significant difference between the AO sensing wavelength ($K$-band) and the science wavelengths, in particular in the 8-13~$\mu$m range. To quantify the wavefront errors induced by water vapor, we analysed $K$-band fringe-tracker data from the GRAVITY instrument at the VLTI\cite{Raghu+26}. This statistical analysis allows us to derive expected wavefront errors due to water vapor seeing at the different representative METIS HCI filters and for different observing conditions\cite{Absil+26}. The results of those analysis are directly used in the simulations presented in this paper. 
\end{itemize}

\begin{figure}
    \includegraphics[width=0.245\linewidth]{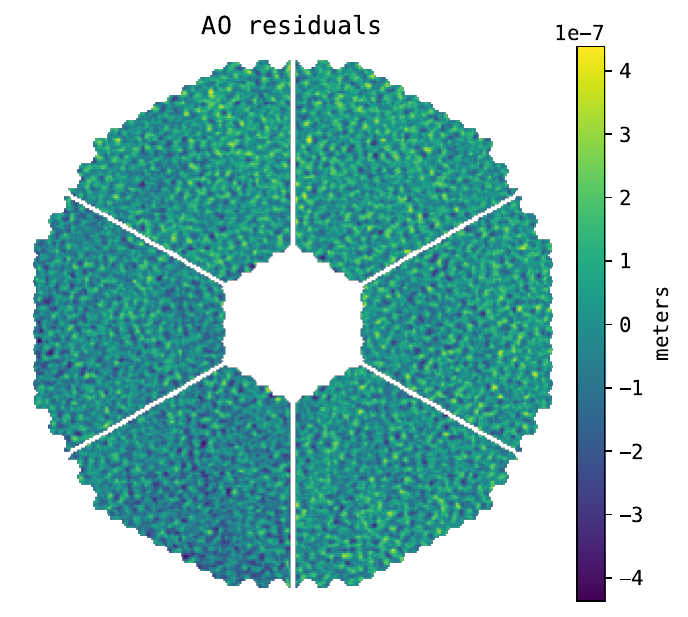}
    \includegraphics[width=0.245\linewidth]{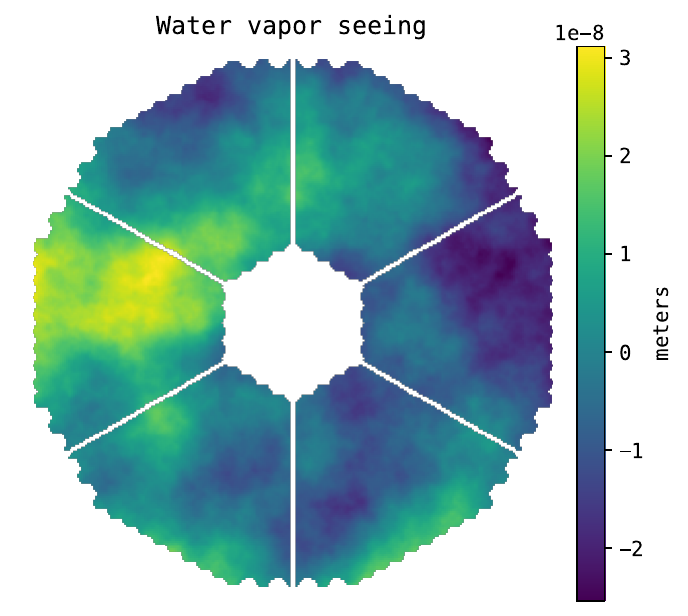}
    \includegraphics[width=0.245\linewidth]{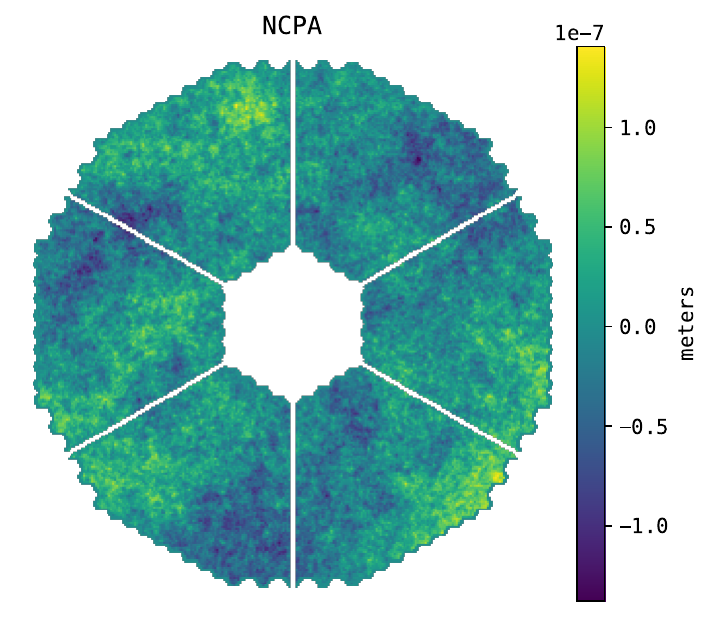}
    \includegraphics[width=0.245\linewidth]{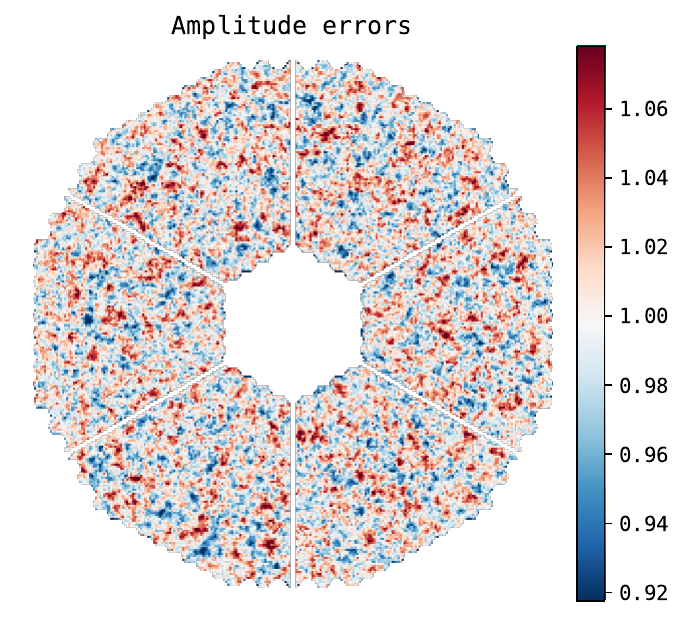}
    \caption{Illustration of the input phase and amplitude errors to our simulations. From left to right: residual atmospheric wavefront error, uncorrected water vapor $L$-band phase error, uncorrected NCPA due to chromatic beam wander, amplitude error due to Talbot effect. }
    \label{fig:error_screens}
\end{figure}

In Fig.~\ref{fig:error_screens}, we illustrate the different phase and amplitude errors. The panels give an impression of the values (in $L$-band) and spatial frequency contents in the different phase and amplitude contributions.
While the amplitude errors cannot be corrected and are used as-is in our simulations, the combined phase errors are corrected by focal-plane wavefront sensing before being fed to HEEPS. This is the topic of the next two sections \ref{sec:fpwfs} and \ref{sec:all together}.

\subsection{Focal plane wavefront sensing}\label{sec:fpwfs}
Considering the major role of  dynamic and differential aberrations (CBW and WV; described in the previous section) on the HCI performance, an instantaneous focal plane wavefront sensing strategy has been developed. Hence, our ability to measure and correct them by offsetting the AO will be a key driver for the final HCI performance.
The baseline focal plane wavefront sensing (FP-WFS) framework for the METIS vortex coronagraphic modes is composed of the Asymmetric Lyot waveFront sensor (ALF)\cite{OrbandeXivry+24} and Quadrant Analysis of Coronagraphic Images for Tip-tilt Sensing (QACITS)\cite{Huby+16}.
%

%QACITS was designed to control and stabilize pointing for vortex coronagraphs and as been deployed on-sky on several instruments, most recently the VLT/ERIS.
QACITS was designed to control and stabilize pointing for vortex coronagraphs. As a small inner working angle coronagraph, the vortex phase mask is highly sensitive to pointing errors, which can degrade performance. QACITS operates by analyzing the quadrant asymmetry of vortex coronagraphic images. The relationship between the measured shift and the tilt (or tip) is non-linear and generally non-bijective when considering the entire image. To address this, QACITS divides the focal plane images in several areas to construct different estimators. The so-called ``outer estimator'', which considers quadrant flux asymmetry in an annulus ranging from 1.7 to 2.3 $\lambda$ /D, offers the largest linear range and sufficient precision. While relatively robust to misalignment and low-order aberrations, it trades off some signal-to-noise ratio (SNR). The linear response of the outer estimator is typically calibrated through a dedicated on-sky procedure (e.g. Ref.~\citenum{OrbandeXivry+24b}), which is also replicated in simulations.

ALF is developed specifically for the METIS vortex coronagraphic modes and targets low-order aberrations beyond tip-tilt. It employs
an asymmetric Lyot stop to encode even pupil aberration modes into observable intensities. This approach is paired with supervised learning to solve the non-linear inversion problem, using science images to obtain phase aberration maps. The Lyot plane was chosen over the entrance pupil for the mask to preserve the coronagraph’s rejection efficiency and minimize the impact of asymmetry on throughput.
For inversion, we currently use deep learning with convolutional neural networks (CNNs) in a supervised
learning framework. Our implementation leverages a modified ResNet-18 architecture, which processes
individual focal plane images as input and outputs modal coefficients describing wavefront
aberrations.
The baseline with ALF is to measure and correct the first 20 modes of a Karhunen-Loève basis adapted to the ELT-M4 mirror\cite{Feldt+24}.

% - Baseline operation
The baseline framerate for ALF and QACITS is 10~Hz to maximize the temporal rejection bandwidth and minimize the temporal error associated with water vapor correction. To extend the operational range to fainter targets, we also analyze 1~Hz operation, trading temporal error for reduced sensor noise (or measurement noise).

% - describe the simulation performed at 10Hz vs 1Hz, L-band vs N-band
To feed our HCI end-to-end simulations, we  determine the operational range and sensing precision of the two sensors as a function of magnitude, filter, and asymmetric Lyot masks.
%
% Through simulations, we analyze  sensor noise to derive operational limits for running the FP-WFS algorithms at 10 Hz (nominal), 1 Hz, and their non-operational threshold. 
% By extension, 
% these operational limits  define  the  operational boundaries of the HCI modes and our simulation grid presented in Section \ref{sec:grid}.

\subsubsection{ALF sensor noise}
For each test case, defined by the four parameters (filter, Lyot mask, magnitude, framerate), we train a dedicated CNN model.  The training of each model is based on datasets of 10,000 entries where we split the data in a 90/10 ratio between training and validation.

To estimate the sensor noise (or measurement noise), we use static non-common path aberrations (NCPAs), distributed across the same number of modes as in training (typically 20). Higher-order aberrations are present in the training datasets but not in the evaluation considered here. Each modal coefficient is randomly sampled from a power-law distribution proportional to $n^{-5/3}$, where $n$ is the radial index. The NCPA amplitude in the training data set is scaled to wavefront errors of 30 nm rms for the $L$ and $M$-bands, or 100 nm rms for the $N1$ and $N2$-bands\footnote{The four bands we refer to here ($L$, $M$, $N1$ $N2$) correspond to specific METIS HCI filters (HCI-L long, CO Ref, N1, N2) with central wavelengths of 3.8, 4.8, 8.7, 11.3 $\mu$m. See  section \ref{sec:grid} for more details.}, which corresponds to the typical WFE regimes expected for the sensors in closed-loop operation. 

The sensor noise is then calculated as the mean residual error (static NCPA minus the ALF-based wavefront correction) from 500 different realizations. This approach does not provide the closed-loop residuals under dynamically changing aberrations. Instead, it characterizes the sensitivity limits of ALF as a function of the previously mentioned parameters: 
\begin{itemize}
    \item Filter: affects PSF sampling, magnitude zero point, and background.
    \item Lyot masks: encode focal plane information differently and have a minor impact on transmission (and thus SNR).
    \item Magnitude and framerate (inverse of exposure time): directly influence the signal-to-noise ratio of the images and are the primary drivers of sensitivity.
\end{itemize}
% filter --which means different PSF sampling, zeropoint and background--, the Lyot masks, the magnitude, the correction framerate.
% Filter, magnitude and framerate($=1/$dit) directly impact the signal-to-noise ratio of the images and are the main driver for the sensitivity. The Lyot masks encode the information in the focal plane in slightly different ways, and have also a minor impact on transmission (thus SNR).
% The baseline framerate for ALF is set to 10~Hz to maximize the temporal rejection bandwidth and minimize the temporal error associated to water vapor seeing correction. To extend the operational range to fainter targets, we also analyze the ALF sensor noise at 1~Hz. This reduced framerate offers a different trade-off between sensor noise  and temporal error. 

% \textit{The results of our simulations, following the methodology described in Section \ref{sec:alf-method}, are presented in Figures \ref{fig:ALF-L} to \ref{fig:ALF-N2} for the L-, M-, N1-, N2-bands, respectively. Each figure shows the ALF sensor noise (wavefront error rms after correction) as a function of magnitude. The results are provided for exposure time of 0.1 seconds (left) and 1 second (right). Each plot includes curves for the different available Lyot masks. Each curve represents the mean value over 500 realizations, with the standard deviation indicated by in semi-transparent areas.
% For details on the Lyot masks, refer to \cite{E-TNT-ULG-MET-1043} and \cite{E-REP-ULG-MET-1033}.}

\begin{figure}
    % \centering
    \includegraphics[width=0.5\linewidth]{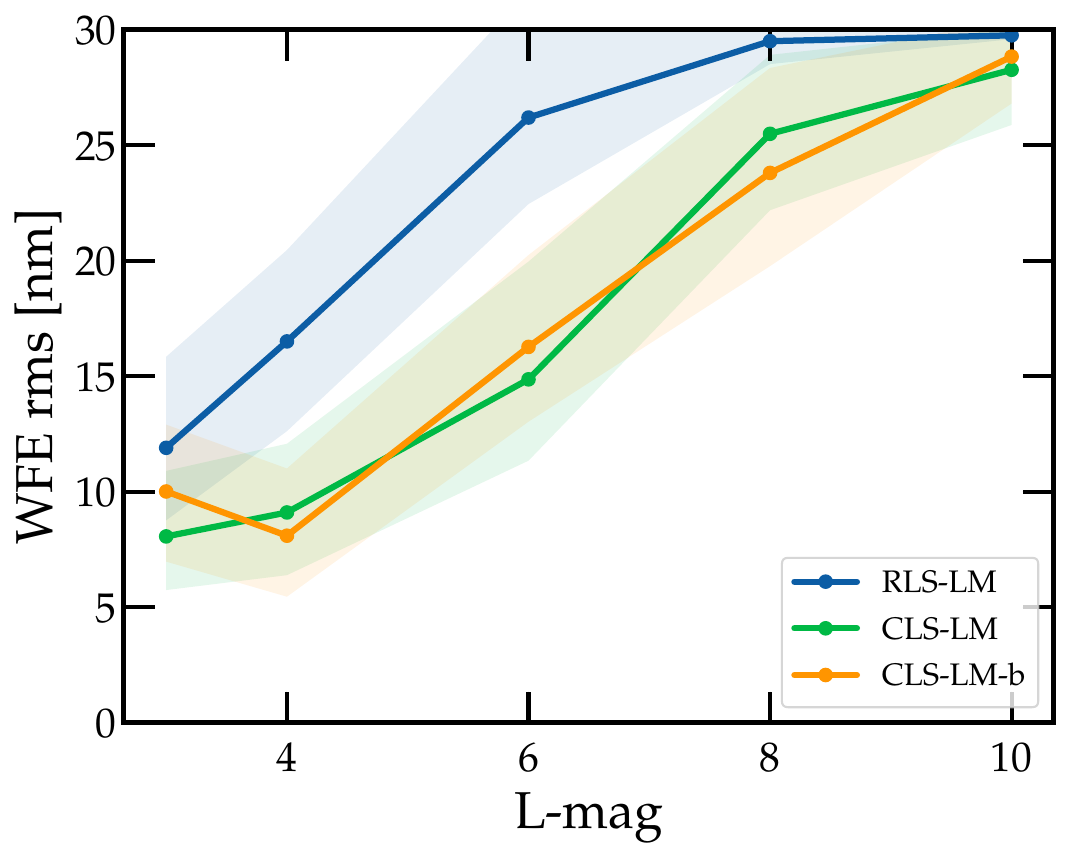}
    \includegraphics[width=0.5\linewidth]{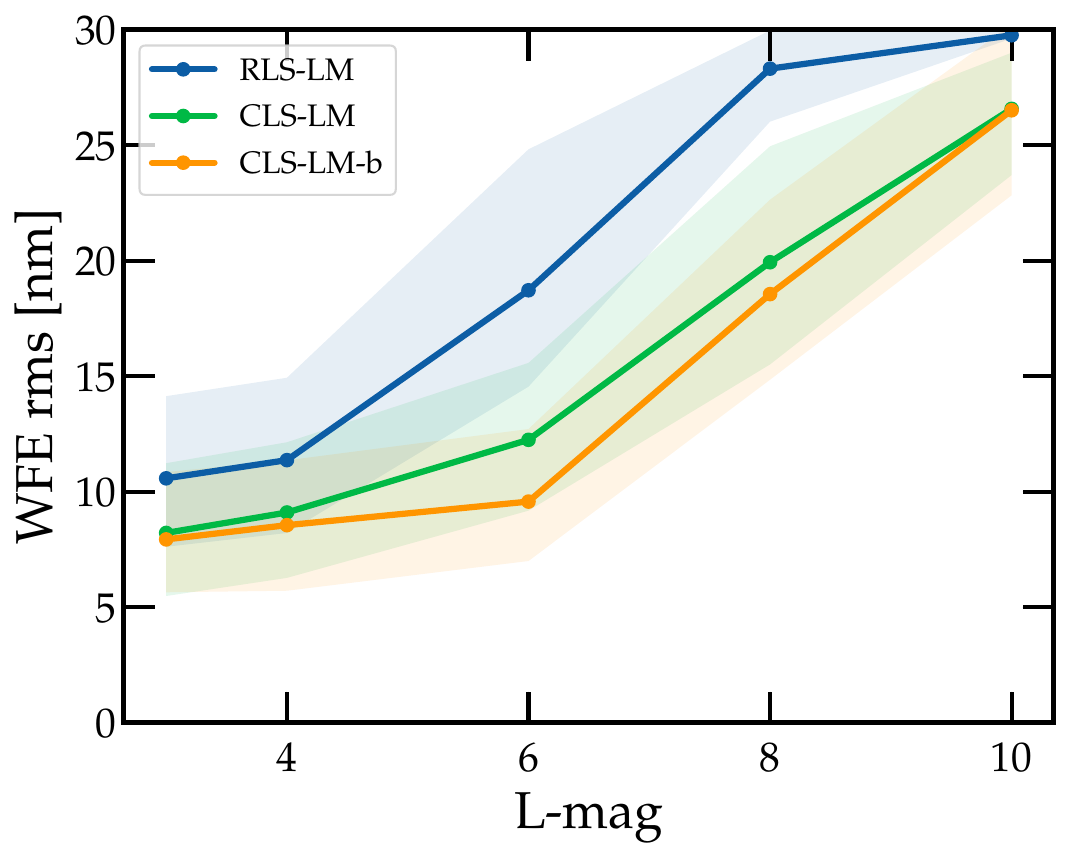}
    \caption{$L$-band ALF sensor noise as a function of magnitude. (Left) for a 10 Hz framerate. (Right) for a 1 Hz framerate. The green and orange curves corresponds to the CVC modes (baseline and backup mask resp.). The blue curve corresponds to the RAVC mode (using the RLS-LM cold stop).}
    \label{fig:ALF-L}
\end{figure}

\begin{figure}
    % \centering
    \includegraphics[width=0.5\linewidth]{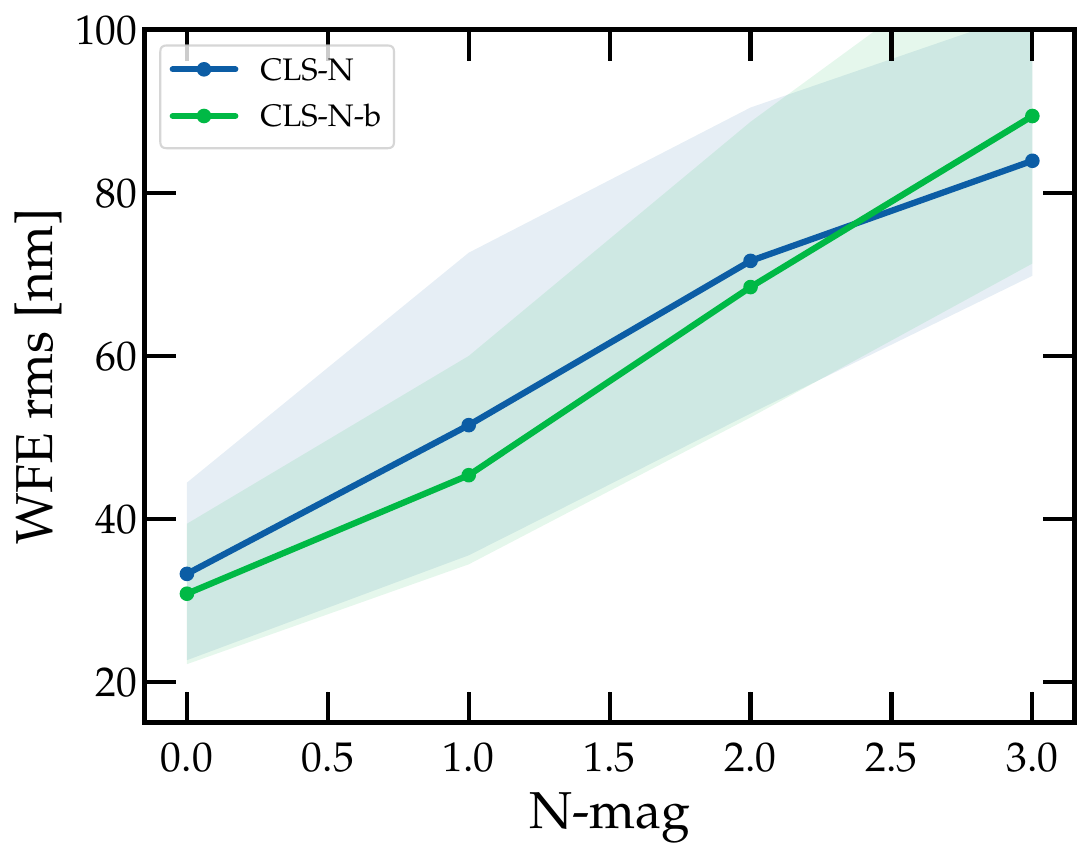}
    \includegraphics[width=0.5\linewidth]{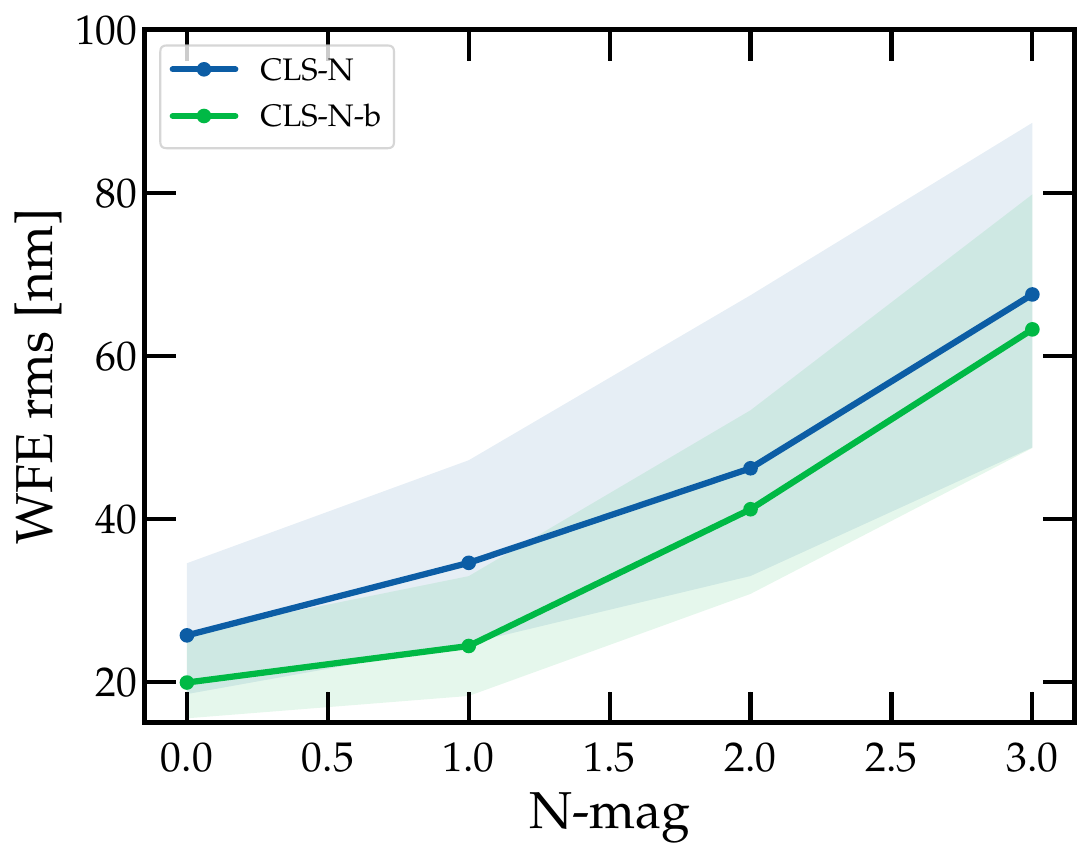}
    \caption{Same as figure \ref{fig:ALF-L} in $N2$-band.}
    \label{fig:ALF-N2}
\end{figure}

The results of our simulations are illustrated in Figs. \ref{fig:ALF-L} and \ref{fig:ALF-N2} for the $L$ and $N2$-bands respectively. Each figure shows the ALF sensor noise (wavefront error rms in the ALF estimation) as a function of magnitude. The results are provided for exposure time of 0.1 second (left) and 1 second (right). Each plot includes curves for the different available Lyot masks. Each curve represents the mean value over 500 realizations, with the standard deviation indicated by  semi-transparent areas.

% - what we learn, and how it is used
In $L$-band, the ALF sensor noise is significantly higher for the RAVC mode, primarily due to its lower throughput.
All the $L$-band curves plateau at approximately 30 nm for faint stars. This plateau is an implicit regularization effect resulting from the CNN training under 30 nm rms. If the training had been performed with higher rms levels, the sensor noise would exceed the 30 nm ceiling.

When increasing the exposure time from 0.1 seconds to 1 second, we observe the expected shift of the sensor noise curve towards fainter magnitudes, by about 1.25 magnitudes (i.e. $2.5 \log{\sqrt{10}}$).

In most cases, particularly in the $N$-band, the conservative masks (denoted as ``-b'') exhibit a slight advantage. Although these masks have reduced throughput, they better encode the phase aberration information in the focal plane images. However, this effect remains minor in most scenarios.

\subsubsection{QACITS sensor noise}
For QACITS, designed for vortex coronagraphy pointing control, we first calibrate the algorithm’s response for a given filter. We then evaluate the pointing error as a function of magnitude.
The calibration process involves incrementally offsetting the on-axis PSF from the vortex center by known
offsets along both the $x$- and $y$-axes. The raw QACITS measurements are recorded, and the response is
determined by performing a linear fit of these measurements against true offset values. The slope of this
linear fit is saved and used subsequently by the algorithm (see for examples Refs.~\citenum{Huby+17,OrbandeXivry+24b}).
%\textbf{Further details and illustrations of the QACITS algorithm calibration can be found in [RD6], Section 5.1.}
Once the QACITS algorithm is calibrated for a specific band, we simulate a series of realistic pointing
offsets to infer the QACITS sensor noise. This is achieved by introducing SCAO residuals and adding
a random pointing error of 1 mas rms, generated from a white power spectral density (PSD) spanning the
0.01 - 0.1 Hz range. The QACITS estimations use the PSF images for a specific integration time (DIT). The residual pointing error is the rms error (estimation minus true value), averaged over tip and tilt.
The true value is the temporal average of the injected pointing error during the integration time.

The QACITS calibration reveals significant model errors (not shown here), with a clear dichotomy between the x- and y-axes and a non-zero offset for a true tip-tilt of zero. We attribute this degradation to the newly designed asymmetric Lyot stops, which slightly deform the post-coronagraphic PSFs, bias the (outer) estimator, and would call for a more refined calibration.
A related effect was observed in a study on Keck\cite{Huby+17}, where the authors characterized the bias induced
by the QACITS outer estimator as a function of the Lyot stop misalignment. This bias can be mitigated by
adjusting the QACITS set point to a non-zero value. While we did not implement such an adjustment in
the present analysis, we emphasize that the model errors dominate the sensor noise at the bright end. This
effect could be mitigated to first order by changing the QACITS set point.

%\textit{An illustration of the QACITS calibration in L-band for the CVC mode is provided in Figure 2-5. This calibration can be compared to the previous results documented in [RD6] at FDR. Notably, the current analysis reveals significantly larger model errors, with a clear dichotomy between the x- and y-axes and a non-zero offset for a true tip-tilt of zero. We attribute this degradation to the newly designed asymmetric Lyot stops, which slightly deform the post-coronagraphic PSFs, bias the (outer) estimator, and would call for a more refined calibration. A related effect was observed in a study on Keck [RD8], where the authors characterized the bias induced by the QACITS outer estimator as a function of the Lyot stop misalignment. This bias can be mitigated by adjusting the QACITS set point to a non-zero value. While we did not implement such an adjustment in the present analysis, we emphasize that the model errors dominate the sensor noise at the bright end. This effect could be mitigated to first order by changing the QACITS set point.}

\begin{figure}[h!]
    \includegraphics[width=0.5\linewidth]{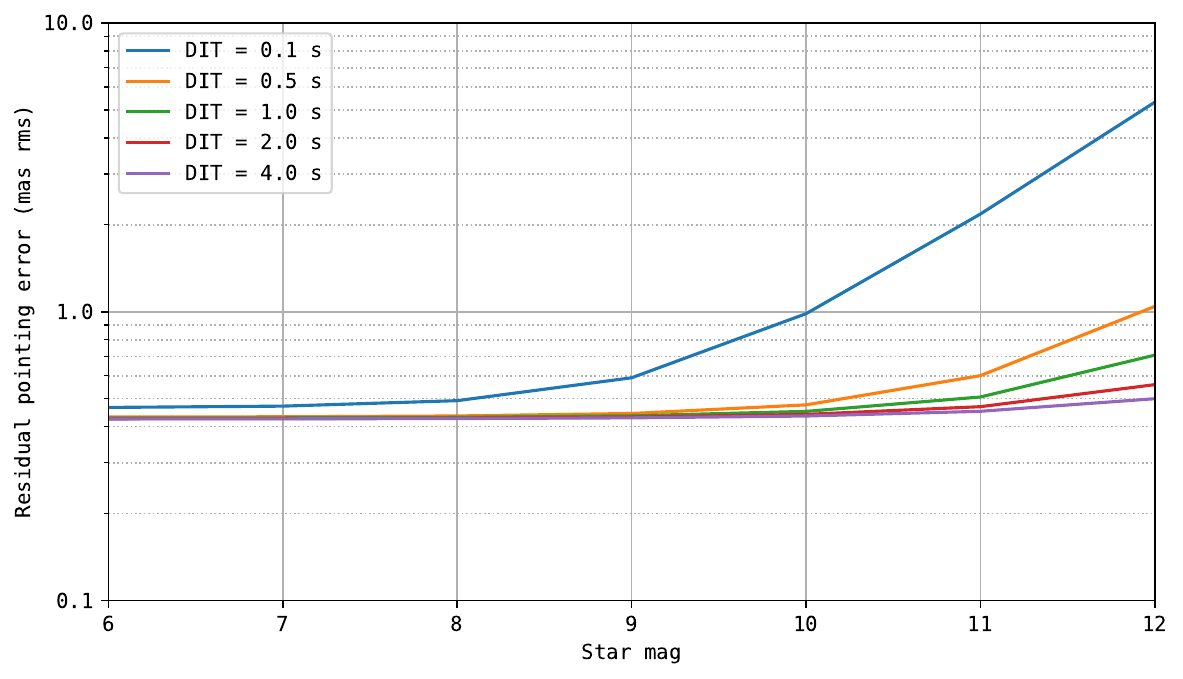}
    \includegraphics[width=0.5\linewidth]{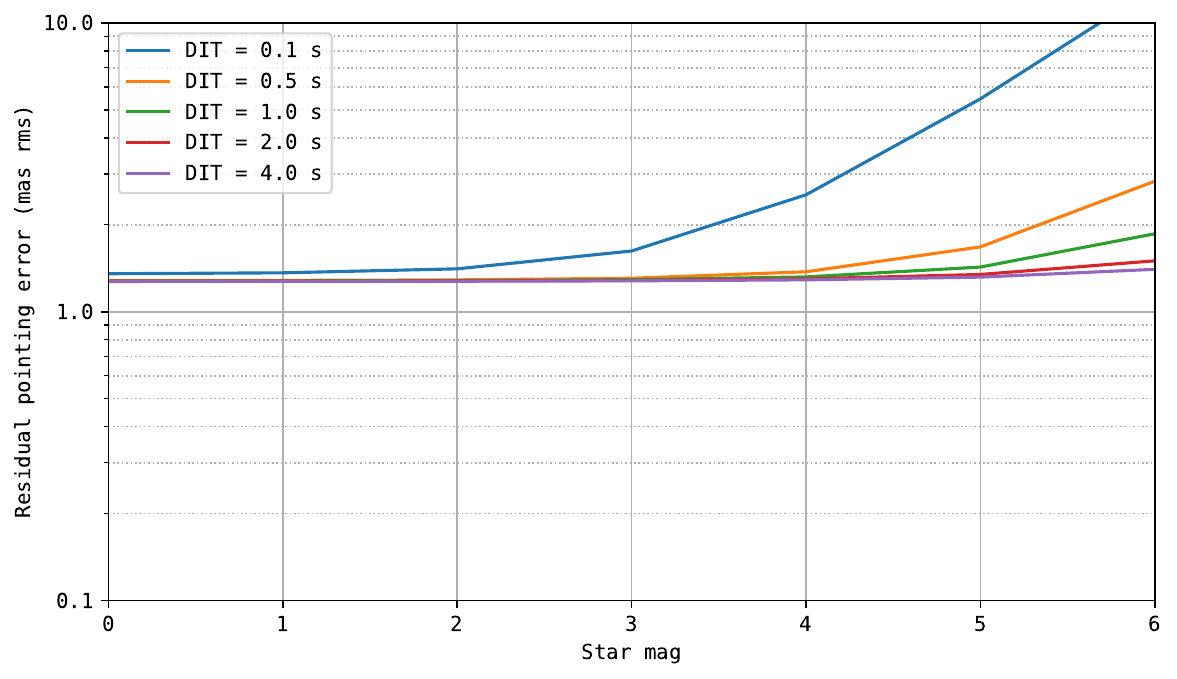}
    \caption{QACITS sensing accuracy in presence of SCAO residuals and for variable pointing errors of 1 mas rms, as a function of the stellar magnitude and integration time. (Left) $L$-band CVC mode. (Right) $N2$-band CVC mode.}
    \label{fig:qacits}
\end{figure}

The results of our pointing error analysis are illustrated in Fig.~\ref{fig:qacits}. It shows that at 10~Hz (0.1~s DIT), we reach a sensing error $\sim$0.5 mas rms for $L \leq 8$, and  $\sim$1.5 mas rms for $N2 \leq 2$.
Since model errors set the floor at the bright end, these sensing errors can be considered as upper limits.

Since QACITS is less sensitive to magnitude variations than the higher-order corrections, the operational limits are primarily driven by ALF. Another key driver for operation limits - discussed in Section \ref{sec:grid} - is the background noise which sets a limit on the efficiency of the coronagraphic mode.

\subsection{Bringing this all together}\label{sec:all together}

Before using the HEEPS pipeline,  we prepare the input phase aberrations as follows:
\begin{enumerate}
    \item We combine the three contributions to the phase aberrations: SCAO residual errors, chromatic beam wander phase errors, and water vapor seeing errors.
    \item We perform closed-loop control of the first 20 Zernike modes at 10~Hz using  focal plane wavefront sensing. The wavefront is obtained directly by projecting the phase screens onto the modal basis, to which we add random modal noise drawn from normal distributions with standard deviations derived from our focal plane WFS analysis, see last section and e.g. Fig.~\ref{fig:ALF-L}.
\end{enumerate}
The resulting residual phase screens obtained from this closed-loop control implementation serves as input  to wavefront propagation and production of instantaneous PSFs in HEEPS. 
For  amplitude aberrations, which cannot be corrected, the errors are directly provided as input to HEEPS without further processing.

\subsection{Case study}
With the  updates described in previous sections, we now evaluate the relative impact of residual phase and amplitude aberrations by selectively including them in our simulations. We perform this analysis for the $L$-band, and $N2$-band where water vapor effects are most pronounced. 
As discussed in section \ref{sec:fpwfs}, METIS  mitigates phase aberrations via focal plane wavefront sensing (FPWFS), while amplitude aberrations cannot be corrected. Thus, the NCPA in our simulations represents residual phase errors after closed-loop control of the first 20 modes at 10~Hz, accounting for a 1-frame delay and realistic FPWFS noise.

\begin{figure}
    \centering
    \includegraphics[width=\linewidth]{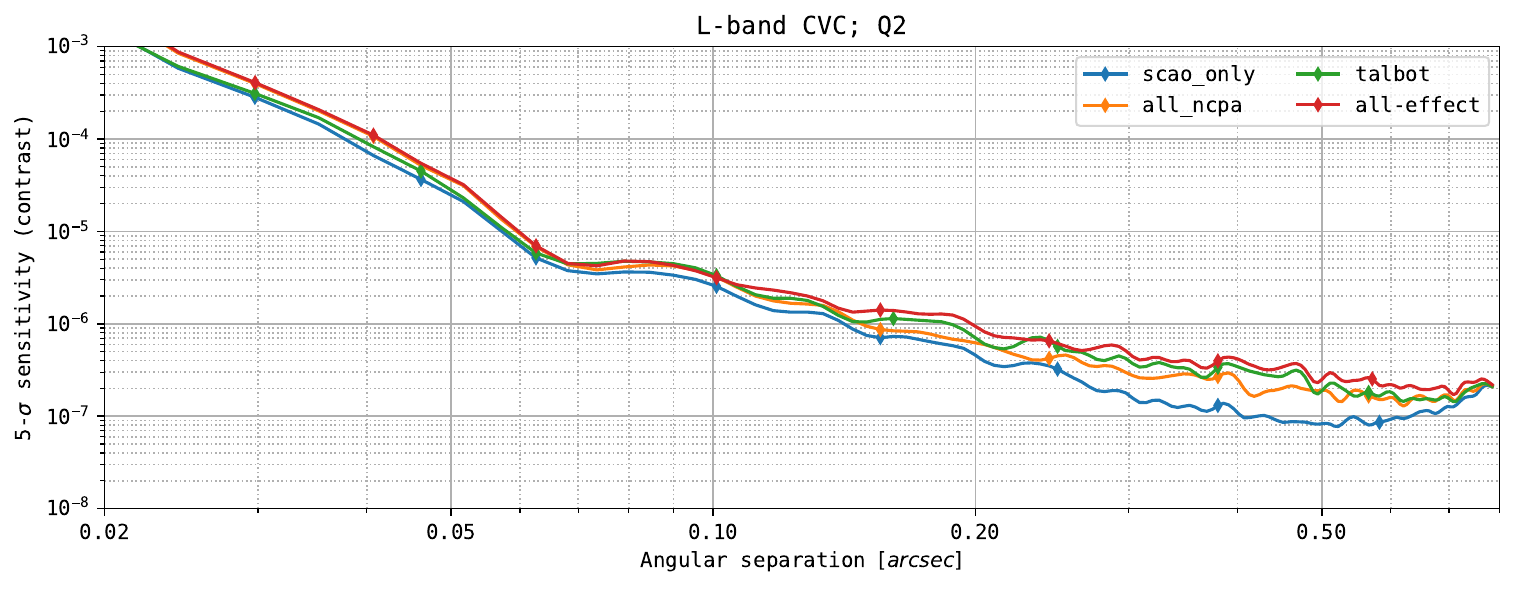}
    \includegraphics[width=1.02\linewidth]{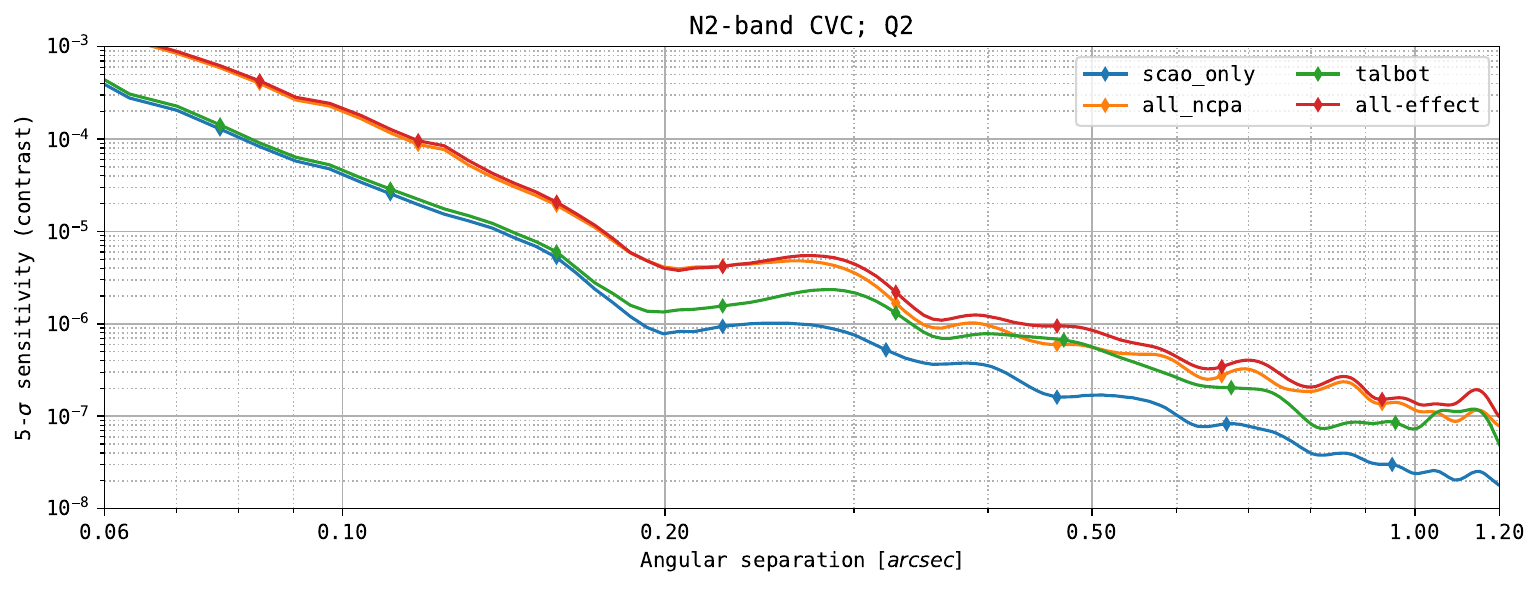}
    \caption{ADI post-processed contrast curves for the CVC $L$-band (top) and CVC $N2$-band (bottom) for various instrumental effects and their combination, but without background noise. The second percentile (\ie~37.5\% of the time conditions are better or equal) is considered for the observing conditions (dry air and water vapor). Individual contrast curves are shown for the dominant error sources: NCPA residuals (CBW and WV) after closed-loop control and Talbot effect for a meridian-centered sequence. The `SCAO only' curve shows the contrast floor set by the adaptive optics residual alone, and the `all-effects' curve combines all three contributions and provides our final performance prediction.}
    \label{fig:test_case}
\end{figure}

The results are presented in Figure \ref{fig:test_case}. 
In $L$-band (Figure \ref{fig:test_case}, top), NCPA residuals and Talbot errors contribute nearly equally to the HCI performance. Talbot errors dominate slightly at mid-frequencies ($>$0.2\arcsec\ to 0.5\arcsec, or $>$10 - 50 $\lambda$/D), causing the combined performance to depart from the `SCAO-only' case in this regime.
In $N2$-band (Figure \ref{fig:test_case}, bottom), the floor set by SCAO residuals is lower due to the inverse wavelength dependence of the wavefront error. However, water vapor seeing  becomes the dominant factor  at small separations ($<$0.35\arcsec) and remains significant up to $\sim$0.5\arcsec. It is worth noting here that the observing conditions (dry air and water vapor seeing) assume the second percentile (\ie~37.5\% of the time conditions are better or equal). Other observing conditions  would lead to slightly different conclusions.
Beyond 0.5 - 0.6\arcsec,  CBW NCPA overtakes over WV NCPA (the CBW/WV breakdown is not shown separately in the figure), while Talbot errors contribute substantially at mid-frequencies ($\sim$0.35\arcsec\ to 0.5\arcsec, or $\sim$6 - 8 $\lambda/D$).

The expected HCI performance, particularly in the $N2$-band, has improved significantly compared to earlier reports (e.g., Refs.~\citenum{Feldt+24,Delacroix+22}). This stems from reduced water vapor-induced wavefront errors\cite{Raghu+26,Absil+26} relative to our earlier estimates, and smaller chromatic beam wander effects. Updates to CBW simulation inputs\cite{Bone+26} reflect that the Common Fore Optics mirrors -- most of which are now produced and qualified -- exhibit surface figure errors better than specified. Additionally, the inverse square power spectral density function previously used to extrapolate low-order errors was conservative. Indeed, the polishing process, using large polishing disks, suppresses mid-spatial frequencies, resulting in  minimal errors between low orders and scratch-dig and roughness levels.

%While those recent updates are favorable to the HCI performance, we will continue to monitor HCI performance as a function of our various system analyses.

% baseline study for L-band and N2-band

%\newpage
\section{SUPPORTING SCIENCE OBSERVATION PLANNING}
\subsection{HCI vortex simulation grid}\label{sec:grid}
As the METIS science team activities ramp up to prepare its GTO program, contrast curves and sequences of coronagraphic PSFs across different stellar magnitudes and wavelengths are needed to support the preparation of future METIS observations, in particular for exoplanets and discs.

To address this demand systematically, we have defined and computed a large grid of METIS HCI simulations. Currently, the grid focuses on the vortex coronagraphic modes (classical vortex and ring-apodized vortex) and explores the following parameter ranges:
\begin{itemize}
    \item the four broadband filters $L$, $M$, $N1$, $N2$ with respective central wavelengths of 3.8, 4.8, 8.7, 11.3~$\mu$m (and bandwidths of $\sim$6\%, 4\%, 13\%, and 21\% respectively). The HEEPS simulations are monochromatic and performed at the central wavelength, but include the derived photometry for each filter (zeropoint and thermal background flux).
    \item three different observing conditions that set the atmospheric turbulence and the water vapor seeing levels (Q1, Q2, Q3, \ie~the 12.5, 37.5, 62.5\% percentiles). 
    In terms of Fried parameter, the Q1, Q2, and Q3 percentiles corresponds to $r_0=$~23.4,  17.8 and 13.9~cm respectively, or seeing values of 0.43\arcsec, 0.57\arcsec, and 0.73\arcsec~at 500~nm. The rms wavefront error due to water vapor seeing in the $N2$-band -- the worst case --  is 102, 147 and 205~nm  for the three quartiles respectively\cite{Raghu+26,Absil+26}. As a comparison, in the $L$-band, WV seeing rms ranges from 8.2 to 16.4~nm rms.
    \item A large magnitude range, from very bright cases at magnitude $-1.5$ (covering targets like $\alpha$~Cen) to the faint end for each filter, where performance is almost exclusively dominated by thermal background (see discussion below). For AO performance, we assume the baseline case with magnitude $K=6$ and the AO correction running at 1~kHz. For fainter targets, we also consider $K=8$ at 1~kHz and $K=9$ at 500~Hz.
\end{itemize} 

The key elements of this simulation grid are summarized in the Table \ref{tab:grid}.

\begin{table} %[ht]
\caption{Key features of the METIS HCI vortex simulation grid.} 
\label{tab:grid}
\begin{center}       
\begin{tabular}{|l|p{70mm}|} 
\hline
\rule[-1ex]{0pt}{3.5ex}  Vortex coronagraphic modes & Classical Vortex Coronagraph (CVC)\\
 & Ring-Apodized Vortex Coronagraph (RAVC)  \\
\hline
\rule[-1ex]{0pt}{3.5ex}  Filters & $L$ (HCI-L long), $M$ (CO ref), $N1$, $N2$  \\
& (3.8, 4.8, 8.7, 11.3~$\mu$m resp.)   \\
\hline
\rule[-1ex]{0pt}{3.5ex}  Magnitude steps & 0.5  \\
\rule[-1ex]{0pt}{3.5ex}   \quad  $L$-band CVC  range & $-1.5$ to 9  \\
\rule[-1ex]{0pt}{3.5ex}   \quad  $L$-band RAVC  range & $-1.5$ to 4  \\
\rule[-1ex]{0pt}{3.5ex}   \quad  $M$-band CVC  range & $-1.5$ to 6  \\
\rule[-1ex]{0pt}{3.5ex}   \quad  $N1$-band CVC  range & $-1.5$ to 4  \\
\rule[-1ex]{0pt}{3.5ex}   \quad  $N2$-band CVC  range & $-1.5$ to 3  \\
\hline
\rule[-1ex]{0pt}{3.5ex}  Observing conditions & Q1, Q2, Q3 percentiles 
    \par (12.5, 37.5, 62.5 \% ) \\
 \hline
\rule[-1ex]{0pt}{3.5ex}  Observing sequence & 1-hr ADI for declination=5\degree 
\par (39\degree~field rotation)\\
\hline
\rule[-1ex]{0pt}{3.5ex}  Key outputs & \par Post-processed ADI contrast curves
\par On-axis PSF cubes 
\par Off-axis PSF reference  \\
\hline 
\rule[-1ex]{0pt}{3.5ex}  Computation details & C\'ECI Lemaitre4 cluster \\
 & 222 simulations \\
 & total of $\sim$ 8880 CPU-hr\\
\hline
\end{tabular}
\end{center}
\end{table}

This grid of simulations was calculated on the  C\'ECI  high-performance computing cluster. Each HCI simulation takes about 40 CPU-hr and requires up to $\sim$200~GB RAM.
The key outputs of this grid are: the post-processed ADI contrast curves, the on-axis PSF cubes, and the reference off-axis PSFs.
The outputs are shared with the METIS science team via two mechanisms: 
\begin{enumerate}
\item an interactive Python notebook distributed via Binder\footnote{https://mybinder.org/} with  a widget that allows direct online exploration of the contrast curve database,
\item on-demand access to the full database ($\sim$3~TB), which also includes the PSF cubes and allows for the generation of more complex mock observations, see section \ref{sec:science}.
\end{enumerate}

% We can now  examine the post-processed contrast curves as a function of magnitude and establish limits where the coronagraphic modes become inefficient -- because they become dominated by thermal background noise -- and normal imaging should rather be used.

\begin{figure}
    \centering
    \includegraphics[width=\linewidth]{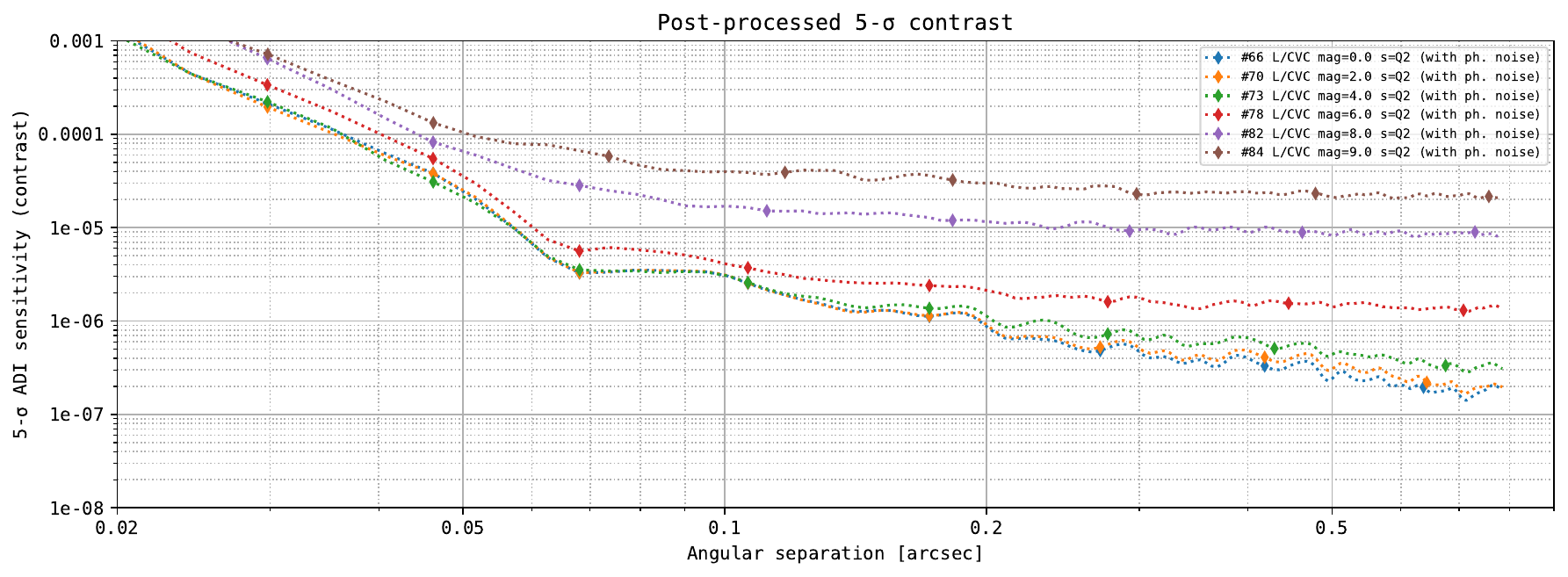}
    \includegraphics[width=\linewidth]{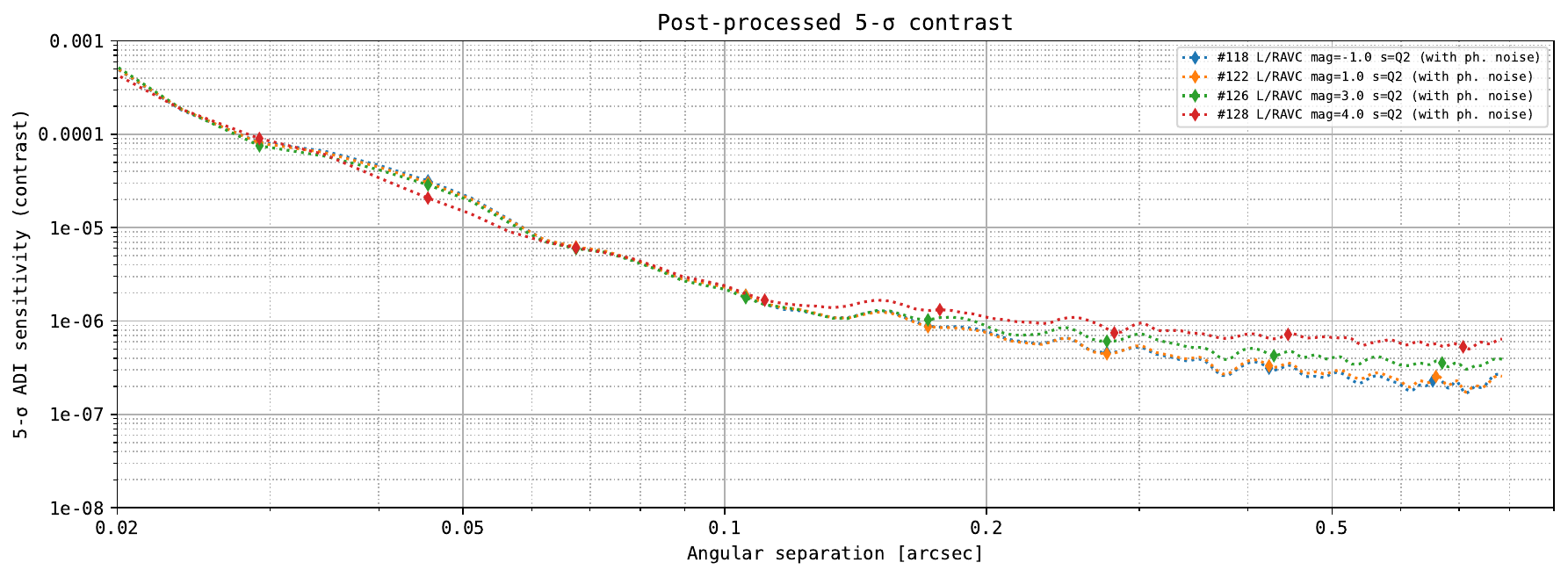}
    \caption{Post-processed ADI contrast curves for different HCI vortex modes and magnitudes. All contrast curves include background and are computed for Q2 percentile observing conditions. (Top) $L$-band CVC mag=[0, 2, 4, 6, 8, 9]. (Bottom) $L$-band RAVC mag=[$-1$, 1, 3, 4].}
    \label{fig:cc_L}
\end{figure}

\begin{figure}
    \centering
    \includegraphics[width=\linewidth]{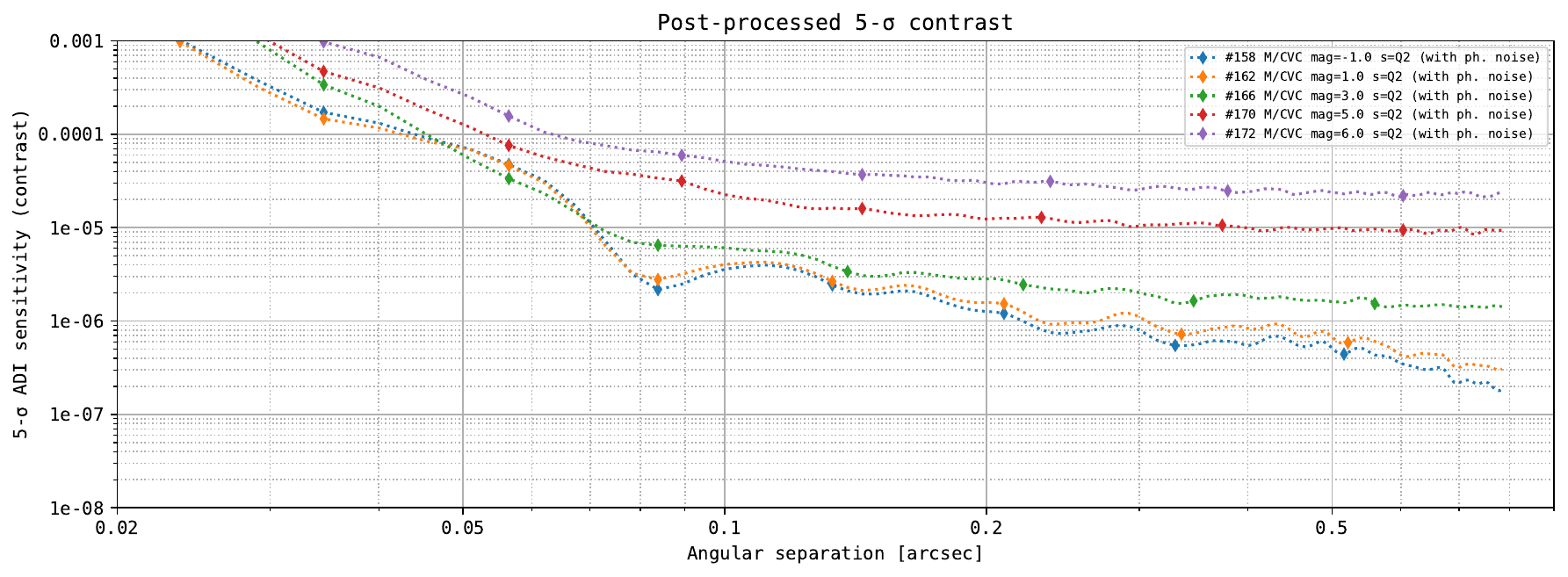}
    \caption{Post-processed ADI contrast curves for $M$-band CVC (Q2 conditions). Magnitudes: $-1$, 1, 3, 5, 6.}
    \label{fig:cc_M}
\end{figure}

\begin{figure}
    \centering
    \includegraphics[width=\linewidth]{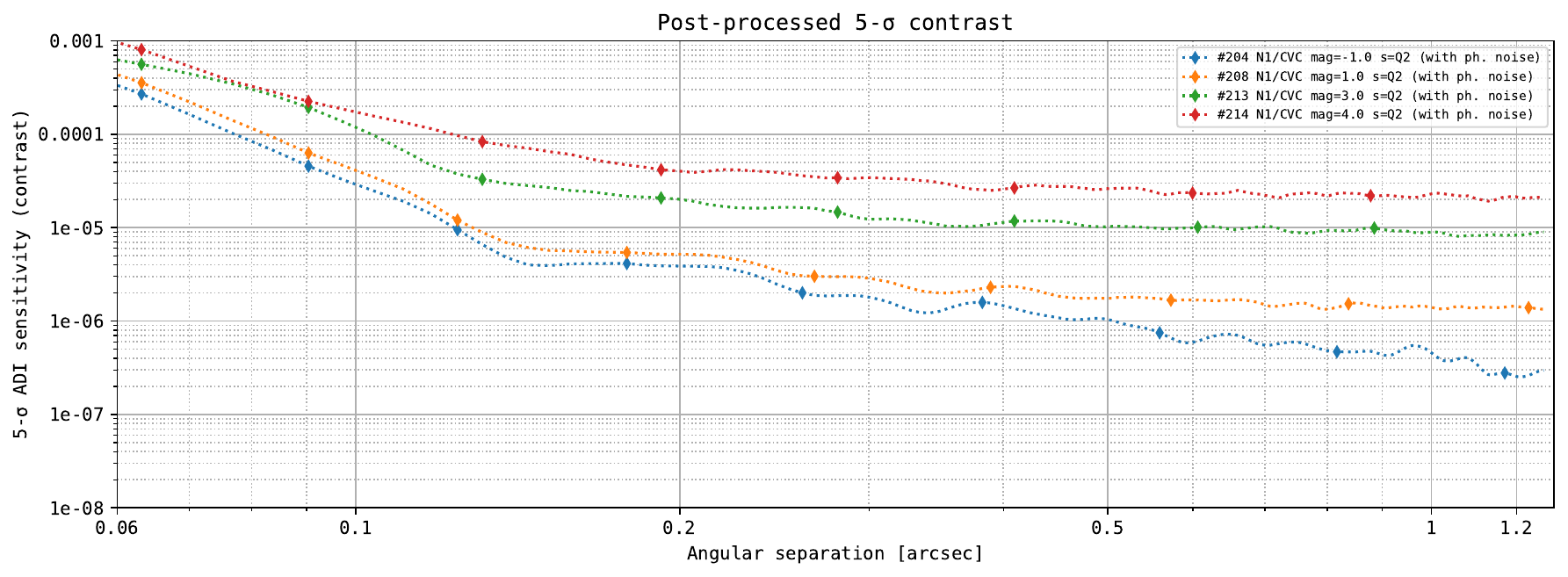}
    \includegraphics[width=\linewidth]{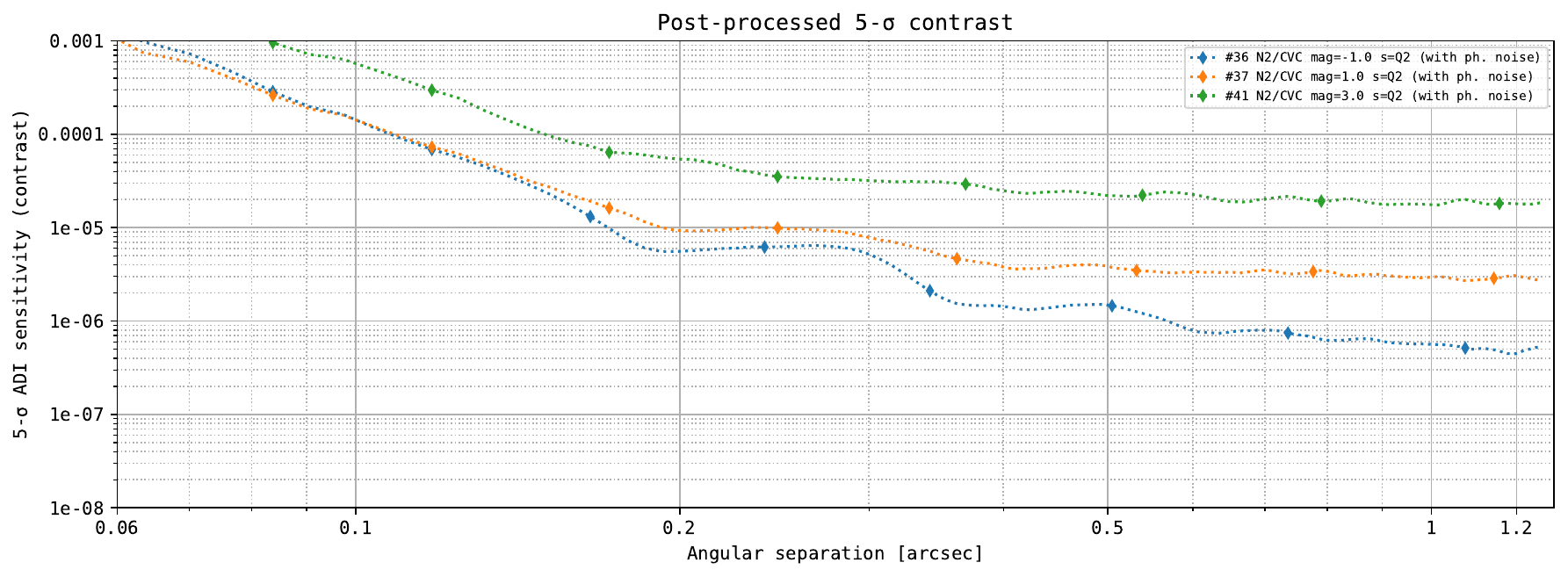}
    \caption{Post-processed ADI contrast curves for (top) $N1$-band CVC and (bottom) $N2$-band CVC (Q2 conditions). $N1$-Magnitudes: $-1$, 1, 3, 4; $N2$-magnitudes: $-1$, 1, 3, 4. }
    \label{fig:cc_N}
\end{figure}

In Figs. \ref{fig:cc_L}, \ref{fig:cc_M} and \ref{fig:cc_N}, we plot a subset of post-processed ADI contrast curves for a range of magnitudes, covering the four filters ($L$, $M$, $N1$, and $N2$ bands) and modes (CVC and RAVC for $L$-band) for Q2 percentile observing conditions. Each plot covers its own specific magnitude range. The post-processed contrast curves can be shared with external people upon request to the authors.

Based on the post-processed contrast curves, we can establish magnitude limits where the coronagraphic modes become inefficient -- because they become dominated by thermal background noise -- and normal imaging should rather be used. For the CVC mode, the fainter magnitude contrast curves are very similar across filters. At these magnitudes, the contrast is almost completely dominated by background noise. Indeed, in these plots, the contrast floor is about $2\times 10^{-5}$ at large separations ($>0.1^{\prime \prime}$ at $L$-band, $>0.2^{\prime \prime}$ at $M$-band, $>0.4^{\prime \prime}$ at $N1$ and $N2$-bands). The contrast degradation at smaller angular separations results from the combined effect of background and reduced transmission due to the vortex rejection, and the FPWFS correction rate dropping from 10~Hz to 1~Hz leading to increased speckle noise. 
%In other words, at the fainter magnitudes plotted in Figs.~\ref{fig:cc_L}, \ref{fig:cc_M} and \ref{fig:cc_N}, the sensitivity limit degrades due to increased photon noise rather than  speckle noise.
For the RAVC mode, the range is narrower with an upper limit defined at $L \sim 4$. Indeed, this mode is optimized for contrast at the expense of throughput. At $L \sim 4$, throughput begins to matter even at small separations ($<0.2^{\prime \prime}$): when comparing the $L$-CVC and $L$-RAVC curves at $L=4$, we find that that the CVC provides a similar or better contrast.

% \hl{Add short discussion on how performance varies from Q1 to Q3...}
The contrast curves show little variation with observing conditions (dry air and water vapor percentiles).
The most significant difference occurs between the Q1 and Q3 observing conditions for the $N2$-band CVC modes at the bright end: in this regime the contrast in the Q3 observing conditions degrades by a factor $<$1.5 at radial separation $<$0.4\arcsec. At larger separations,  contrast is essentially the same and dominated by uncorrected phase and amplitude errors (the phase spatial frequency at those angular separations are beyond the control space provided by the FPWFS).

% \newpage
\subsection{A showcase example: observation of a protoplanetary system with METIS}\label{sec:science}

Beyond the assessment of detection limits using the grid of post-processed ADI contrast curves, our simulated datasets can be used to assess the benefit of  more advanced post-processing algorithms and predict the observations of extended objects, in particular protoplanetary systems featuring substructures such as spirals and gaps.

\begin{figure}[h!]
    \centering
    \includegraphics[width=\linewidth]{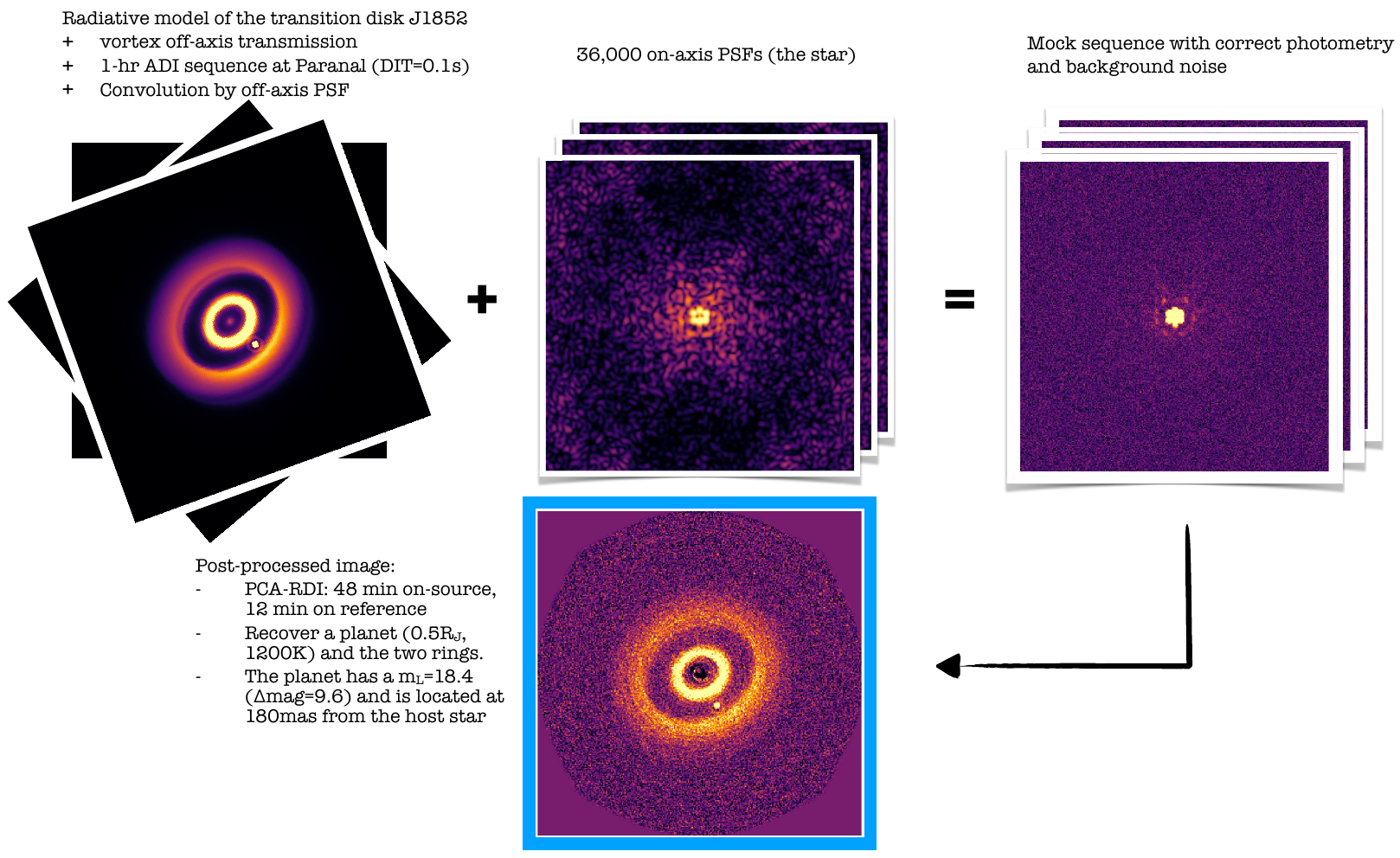}
    \caption{Flowchart of METIS protoplanetary system modelling pipeline, comprising seven key steps: disc model, vortex transmission, ADI sequence generation, PSF convolution, on-axis PSF addition, photometry, and post-processing. The RXJ1852 system is showns as an illustrative example.}
    \label{fig:protoplanet}
\end{figure}

Based on a synthetic disc model of a protoplanetary system (obtained, e.g., by combining smoothed particle hydrodynamics simulations\cite{Price+18} with radiative transfer calculations\cite{Pinte+06} to compute the mid-infrared photometry), we can generate realistic mock METIS observations, which can then be used to predict and assess the recovery of the circumstellar structure.
This advanced modelling approach works as follows:
\begin{itemize}
\item Apply the off-axis vortex transmission curve to the synthetic model,
\item Generate the 1-hr ADI observing sequence,
\item Convolve the circumstellar scene with the off-axis PSF,
\item Add the on-axis coronagraphic PSFs with the appropriate scaling to represent the host star,
\item Apply the appropriate photometry with stellar and background photon noise.
\end{itemize}
The generated sequence can then be post-processed using differential imaging, such as PCA-RDI\footnote{Principle Component Analysis leveraging the Reference Differential Imaging strategy, where the reference data is used to obtain a model of the PSF for subtraction.}.
This modelling and post-processing  pipeline is illustrated in a flowchart in Fig.~\ref{fig:protoplanet}, showcasing the predicted $L$-band observation  of a protoplanetary system RXJ1852\cite{Villenave+19}. 
% \hl{Explain why RXJ1852 is a nice showcase example.}
The host star has a magnitude $L=8.8$ and is observed in the $L$-band CVC mode, assuming Q2 observing conditions.
The 1-hr observing sequence is split into 48-min on-source and 12-min on-reference observations. The mock observations are processed via PCA-RDI (using 1 principal component) and reveal a hypothetical exoplanet responsible for the cavity in the transition disk. The planet has a magnitude $L = 18.4$ ($\Delta L = 9.6$) and is located at 180~mas ($\sim$9~$\lambda$/D) from the host star. This simulation, while already fairly detailed, remains optimistic as it assumes an ideal reference star of the same magnitude and observed under similar conditions.

While a detailed analysis of this observing sequence is beyond the scope of this paper, the recovery of the companion and disc/gap structure with high significance  showcases the METIS capabilities.
%, but we can highlight that the companion is recovered with high significance, as well as the disc/gap structure.
Indeed, although not particularly challenging for METIS (e.g., the planet is located at a relatively large separation), this  example demonstrates how its spatial resolution, sensitivity, and wavelength range will enable the discovery of low-mass embedded planets in such cavities, thereby advancing our understanding of substructures and their driving processes. The modelling strategy and the prediction of protoplanets observation with METIS will be the subject of a forthcoming paper (Hammond et al. in prep\cite{Hammond2026}).

\section{Conclusion \& future work}
This proceeding presents updates to the METIS instrument model and a large grid of HCI vortex simulations across filters and magnitudes, allowing us to define operational regimes of the HCI modes and enabling performance predictions by the science team for, e.g., exoplanet yield or proto-planetary disk observations.

Looking ahead, we will extend this grid to non-coronagraphic imaging, including cases with and without focal plane wavefront sensing, fainter magnitudes, and accounting for AO limits and detector saturation. This will enable a complete assessment of METIS high-contrast imaging capabilities from bright to faint targets across multiple filters.

\appendix    %>>>> this command starts appendixes
% \section{MISCELLANEOUS FORMATTING DETAILS}

\acknowledgments % equivalent to \section*{ACKNOWLEDGMENTS}  

\noindent\textit{Facility:} Some of the required computational resources were provided by the Consortium des Équipements de Calcul Intensif (CÉCI), funded by the Fonds de la Recherche Scientifique de Belgique (F.R.S.-FNRS) under Grant No 2.5020.11, and by the Walloon Region.

\noindent\textit{Software:} This work makes use of the Python programming language, in particular packages including \texttt{HEEPS} (Refs.~\citenum{Carlomagno+20,Delacroix+22} and this proceeding), \texttt{PROPER}\cite{Krist07}, \texttt{vip\_hci}\cite{Christiaens+23}, \texttt{HCIPy}\cite{Por+18}.

\noindent\textit{AI tools:}  Vibe (Mistral AI) was used for light language editing and proofreading. The interactive widget to explore the contrast curve database was developed with the assistance of GitHub Copilot (Claude Sonnet 4.6).

% References
\bibliography{report} % bibliography data in report.bib
\bibliographystyle{spiebib} % makes bibtex use spiebib.bst

\end{document}